\documentclass[twocolumn]{aa}
\usepackage{graphicx}
\usepackage{txfonts}

\newcommand{\myrule}{\rule[-0.1cm]{0.cm}{0.7cm}} 

\begin{document}
   \title{Radial velocity survey for planets and brown dwarf companions \\
to very young brown dwarfs and very low-mass stars in Cha\,I\\
with UVES at the VLT\thanks{Based on observations 
obtained at the Very Large Telescope of the 
European Southern Observatory at Paranal, Chile 
in program 65.L-0629, 65.I-0011, 268.D-5746, 72.C-0653.}
         }

   \titlerunning{Radial velocity survey of brown dwarfs and very low-mass stars in Cha\,I
with UVES at the VLT}

   \subtitle{}

   \author{V. Joergens
          \inst{1}
          }

   \offprints{V. Joergens, \email{viki@strw.leidenuniv.nl}}

   \institute{Leiden Observatory / Sterrewacht Leiden,
              P.O.Box 9513, 2300 RA Leiden, Netherlands
}

   \date{Received 11 May 2005; accepted 6 Sept. 2005}

   \abstract{We present results of a radial velocity (RV) survey for planets and 
brown dwarf (BD) companions to very young BDs and (very) low-mass stars
in the Cha\,I star-forming cloud. 
Time-resolved high-resolution echelle spectra 
of Cha\,H$\alpha$\,1--8 and Cha\,H$\alpha$\,12 (M6--M8), B34 (M5), CHXR\,74 (M4.5), 
and Sz\,23 (M2.5) were taken with UVES at the VLT between 2000 and 2004.
The precision achieved for the relative RVs range between 40 and 670\,m\,s$^{-1}$ and is 
sufficient to detect Jupiter mass planets around the targets. This is the first
RV survey of very young BDs. It probes multiplicity, which is a key parameter
for formation in an as yet unexplored domain, in terms of age, mass, and orbital separation. 
We find that the subsample of ten BDs and 
very low-mass stars (VLMSs, M$\la$0.12\,M$_{\odot}$, spectral types M5--M8)
has constant RVs on time scales of 40 days and less. 
For this group, estimates of upper limits for masses
of hypothetical companions range between 0.1\,M$_\mathrm{Jup}$ and 
1.5\,M$_\mathrm{Jup}$ for assumed orbital separations of 0.1\,AU.
This hints at a rather small multiplicity fraction for very young BDs/VLMSs, for 
orbital separations of $\la$0.1\,AU and orbital periods of $\la$40 days.
Furthermore, the non-variable objects  
demonstrate the lack of any significant RV noise due to stellar activity
down to the precision necessary to detect giant planets.
Thus, very young BDs/VLMSs are suitable targets for 
RV surveys for planets.
Three objects of the sample exhibit significant RV variations with peak-to-peak RV differences of 
2--3\,km\,s$^{-1}$. For the highest mass object observed with UVES (Sz\,23, $\sim$0.3\,M$_{\odot}$),
the variations are on time scales of days, which might be explained by rotational modulation.
On the other hand, the BD candidate Cha\,H$\alpha$\,8 (M6.5) and the low-mass star CHXR\,74 
(M4.5) both display significant RV variations on times scales of $\ga$ 150 days, while 
they are both RV constant or show only much smaller amplitude variations
on time scales of days to weeks, i.e. of the rotation periods.
A suggested explanation for the detected RV variations of CHXR\,74 and 
Cha\,H$\alpha$\,8 is that they are caused by giant planets or very low-mass BDs of 
at least a few Jupiter masses orbiting with periods of several months or longer.
Thus, the presented RV data indicate that orbital periods of companions to very young BDs 
and (very) low-mass stars are possibly 
several months or longer, and that orbital separations are
$\ga$ 0.2\,AU.
This parameter range has not been covered for all targets yet, but will 
be probed by follow-up observations.
Furthermore, we show that 
the scaled down equivalent to the BD desert found around solar-like stars would be 
a \emph{giant planet desert} around BD and VLMS primaries, if formed
by the same mechanism. The present data test its existence for the targets
in the limited separation range of the survey. So far, 
no hints of companions in a `giant planet desert' have been found.

   }

   \maketitle
%

\section{Introduction}

In the last ten years, more than 150 extrasolar planets have been detected by
radial velocity (RV) surveys 
(e.g. Moutou et al. 2005, Marcy et al. 2005a for recent discoveries).
The RV technique traces periodic RV variabilities caused by the wobble of the 
primary object induced by an orbiting mass. Other sources of RV variability, 
like surface activity, can mimick a companion. 
Therefore, up to now RV surveys for planets have been restricted to considerably old primaries with 
ages on the order of a few billion years and with mainly solar-like spectral types. 
The youngest RV planet known to date is orbiting
the zero-age main sequence star $\iota$\,Hor with an estimated age in
the range of 30\,Myr to 2\,Gyr (K\"urster et al. 2000).
Very recently, evidence of substellar companions, possibly of planetary mass,
around the young BD 2M1207\,A (age $\sim$5--12\,Myr, Chauvin et al. 2004, 2005) 
and around the very young star GQ~Lup (age $\sim$0.1--2\,Myr, Neuh\"auser et al. 2005) 
has been found from direct AO imaging.
Furthermore, most planets known to date orbit around solar-mass stars with spectral
types of late-F, G, and early-K with the exception of two planets around the 
M4-dwarf Gl\,876 (Delfosse et al. 1998, Marcy et al. 1998, 2001), a planet around 
the M2.5 dwarf GJ\,436 (Santos et al. 2004, Butler et al. 2004) and the 2M1207 system (see above).

Further progress in the field of extrasolar planets is expected from the search for planets 
around very young, as well as very low-mass primaries. The
detection of \emph{young} planets and a census of planets around
stars of all spectral types and even around BDs, is an important step
towards understanding planet formation. 

Detection of planets around BDs, as well as of
young BD binaries (BD--BD pairs),
would, on the other hand, constrain the formation of
BDs (see Joergens 2005a for a recent review of BD formation).
Besides
indications of a possible planetary mass object in orbit around 2M1207
(Chauvin et al. 2005), no planet of a BD has been found yet.
In recent years, several BD binaries have been detected in the field, 
mainly by direct imaging (e.g. Mart\'{\i}n et\,al.\,1999, 2000, Koerner et\,al.\,1999,
Reid et al. 2001, Lane et al. 2001, Kenworthy et al. 2001, Close et al. 2002a, 2002b,
Bouy et al. 2003) and about three by spectroscopic surveys 
(Basri \& Mart\'{\i}n 1999, Guenther \& Wuchterl 2003).
These detections of companions 
to nearby field BDs give important insight into the 
substellar binary population at ages of a few billion years.
However, these results represent only a boundary condition, 
which is not necessarily matched at any earlier time.
Therefore, it is useful for the current discussion of 
BD formation scenarios to study multiplicity at \emph{very young} ages.
However, besides
indications from direct imaging 
for the binarity of Cha\,H$\alpha$\,2 (Neuh\"auser et al. 2002)
and 2M\,1101-7732 (Luhman 2004) in the Cha\,I cloud and of DENIS-P\,J18590.9-370632 in the R-CrA 
star-forming region (Bouy et al. 2004), all other known BD binaries are fairly old. 

In order to probe both BD multiplicity at a very young age and the occurrence of planets
around very young and very low-mass (substellar) primaries, 
we initiated an RV survey in the Cha\,I star-forming cloud 
with the UV-Visual Echelle Spectrograph (UVES) at the Very Large Telescope (VLT).
The targets are very young BDs and (very) low-mass stars in the center of Cha\,I
at an age of only a few million years.
This is the first systematic RV survey for companions 
around young BDs and VLMSs.
We present evidence in this paper that they show only very small 
amplitude RV variability due to activity and that they are suitable 
targets for RV planet surveys.
This opens up a new parameter range for the search for extrasolar planets, namely
looking for very young planets at very close separations. 
Thus, the initiated RV survey studies the existence of companions in what is an as yet unexplored
domain, not only in terms of primary masses (substellar regime)
and ages (a few million years), but also in terms of companion masses
(sensitive down to planetary masses) and separations (smaller than about 2\,AU). 
It might sample a substantially different companion formation mechanism than 
the one represented by BD binaries detected so far by direct imaging.

First results were obtained by Joergens \& Guenther (2001) 
within the framework of this survey based on UVES spectra taken in 2000
on mean RVs, projected rotational velocities $v \sin i$, and Lithium absorption,
as well as on the kinematics of the BD population in Cha\,I
in comparison with that of T~Tauri stars in the same region. 
The data analysis was improved, and revised RVs were then measured by Joergens (2003).
Additional UVES spectra were taken in 2002 and 2004. 
In the paper on hand, the time-resolved RVs measured from 
all UVES spectra taken during this survey between 2000 and 2004
are finally presented and analysed in terms of  
a search for spectroscopic companions down to planetary masses.
This enlargement of the data set and the improved data reduction allowed an
improved kinematic study of very young BDs based on the mean RVs measured 
with UVES, published elsewhere (Joergens 2005b).

The paper is organized as follows:
Sect.\,\ref{sect:sample} introduces the observed sample of BDs and (very) low-mass stars
in Cha\,I. In Sects.\,\ref{sect:spec} and \ref{sect:rvs},
the acquisition and reduction of 
high-resolution UVES spectra and the measurement of RVs are described. 
In Sect.\,\ref{sect:results}, the results are presented and discussed.
Finally, Sect.\,\ref{sect:concl} contains conclusions and a summary.

\section{Sample}
\label{sect:sample}

The targets of this RV survey are BDs and (very) low-mass stars 
with an age of a few million years situated in the center of the 
nearby ($\sim$160\,pc) Cha\,I star-forming cloud 
(Comer\'on, Rieke \& Neuh\"auser 1999; Comer\'on, Neuh\"auser \& Kaas 2000, 
Neuh\"auser \& Comer\'on 1998, 1999).
Membership in the Cha\,I cluster and, therefore, 
the youth of the objects, is well established based on 
H$\alpha$ emission, Lithium absorption, spectral types, and RVs 
(see references above, Joergens \& Guenther 2001, Joergens 2005b).

UVES spectroscopy has been performed so far for 
Cha\,H$\alpha$\,1--8 and Cha\,H$\alpha$\,12, B34, CHXR\,74, and Sz\,23.
Two of them (Cha\,H$\alpha$\,1, 7) are classified as bona fide BDs 
(M7.5--M8) with mass estimates of 30--40\,M$_{\odot}$, 
five (Cha\,H$\alpha$\,2, 3, 6, 8, 12) as BD candidates 
(M6.5--M7) with mass estimates of 50--70\,M$_{\odot}$,
two (Cha\,H$\alpha$\,4, 5) as 
VLMSs (M6) with masses close to the substellar borderline ($\sim$0.1\,M$_{\odot}$),
one (B\,34) as VLMS (M5) with a mass estimate of 0.12\,M$_{\odot}$,
and two (CHXR\,74, Sz\,23) as low-mass T~Tauri stars (M4.5, M2.5) with 
0.17\,M$_{\odot}$ and 0.3\,M$_{\odot}$, resp.

\section{Acquisition and reduction of UVES spectra}
\label{sect:spec}

High-resolution spectra have been taken so far for twelve BDs and 
(very) low-mass stars in Cha\,I between the years 2000 and 2004 with the cross-dispersed 
echelle spectrograph UVES (Dekker\,et\,al.\,2000) attached to the 8.2\,m Kueyen telescope of the 
VLT operated by the European Southern Observatory at Paranal, Chile.
For each object, at least two spectra separated by a few weeks have been obtained
in order to monitor time dependence of the RVs.
For several objects, more than two and up to twelve spectra were taken. 

The observations were performed with the red arm of the two-armed UVES 
spectrograph equipped with a mosaic of two CCDs. The mosaic is made of
a 2K\,$\times$\,4K EEV chip (pixel size 15\,$\mu$m) for the 
blue part of the red arm and an MIT-LL CCD 
for the red part of the red arm.
The wavelength regime from 6600\,{\AA} to 10400\,{\AA} was covered
with a spectral resolution of $\rm \lambda / \Delta \lambda=40\,000$. 
A slit of  1$^{\prime\prime}$ to 1.2$^{\prime\prime}$ was used.

A standard CCD reduction of the two-dimensional UVES echelle frames, 
including bias correction, flat fielding and cosmic ray elimination, 
was performed with
IRAF\footnote{IRAF is distributed by the National Optical
   Astronomy Observatories,
   which is operated by the Association of Universities for Research in
   Astronomy, Inc. (AURA) under cooperative agreement with the National
   Science Foundation.}.
The flat field correction of the small-scale pixel-to-pixel variations of the 
CCD was performed with a master flat field frame
created by taking the median of several flat field exposures. 
The master flat was normalized, in order to remove large-scale structures 
by fitting its intensity along the 
dispersion by a third order fit, while setting all points outside the
order aperture to 1 and dividing the master flat by this fit.

Subsequently, one-dimensional spectra were extracted,
including a correction for sky background light.
Compared to the previous reduction of UVES spectra by Joergens \& Guenther (2001),
this reduction step was improved, as in Guenther \& Wuchterl (2003), 
by extracting first each echelle order as 
a two-dimensional spectrum, then performing the sky subtraction on these 2D frames,
and finally extracting the one-dimensional science spectrum.
No rebinning was done in order to achieve a high RV precision.
%

Finally, the spectra were wavelength calibrated using the echelle package 
of IRAF. This was done in a first step by the use of Thorium-Argon arc
spectra. In order to achieve a high wavelength and therefore RV precision, 
an additional correction by means of 
telluric O$_2$ lines (B-band centered at 6880\,{\AA}) produced in the Earth's atmosphere
was applied.
It has been shown that they are stable up to 
$\sim$\,20\,m\,s$^{-1}$ (Balthasar et al. 1982, Caccin et al. 1985).

\section{Radial velocities}
\label{sect:rvs}

RVs were determined by a cross-correlation of plenty of 
stellar lines of the object spectra against a template spectrum and locating 
the correlation maximum.
For measuring Doppler shifts of stellar features, appropriate wavelength regions 
were selected that are not affected by telluric lines, cosmetic defects of the CCD, or 
fringes of the CCD in the near-IR.
A heliocentric correction was applied to the observed RVs.
In several cases, the RV derived for one night was based on
two consecutive single spectra to provide two independent measurements of the RV.
This allows a solid estimation of the error of the relative RVs based on the standard deviation
for two such data points.
Tables\,\ref{tab:bds} and \ref{tab:tts} list the resulting 
heliocentric RVs, error estimates, and a mean RV for each target.
RV values based on two consecutive single spectra obtained in the same 
night are marked 
with an asterisk in the last column of these tables.
These error measurements depend linearly, as expected, on the signal-to-noise (S/N) of the spectra.
This linear relationship is used in turn to estimate errors for RV data points,
which are based on only one measurement per night.
An RV precision between 40\,m\,s$^{-1}$ and 
670\,m\,s$^{-1}$, depending on the S/N of the individual spectra, was achieved
(last column of Tables\,\ref{tab:bds} and \ref{tab:tts}).
We note that the precision of the RVs is limited by 
the S/N of the spectra and not by systematic effects. 
The relatively high precision that was achieved for the \emph{relative} 
velocities does not apply to the absolute velocities due to additional
uncertainties in the zero point. 
A mean UVES spectrum of selected high S/N spectra of the very low-mass 
M6-type star Cha\,H$\alpha$\,4 served as a template. 
The zero point of the velocity was determined based on a fit to the blend of the prominent 
lithium lines at $\lambda\lambda$\,6707.76 and 6707.93\,{\AA} in three different high S/N spectra.
The standard deviation of these fits of 400\,m\,s$^{-1}$ was assumed as 
an additional error for the absolute velocities.
An observing log listing all individual measured RVs is given in 
Tables\,\ref{obslog1} and \ref{obslog2} in the appendix.

\section{Results}
\label{sect:results}

The monitored RVs are constant within the measurement errors for the majority of the 
observed BDs and (very) low-mass stars, whereas for three of the targets, 
our observations reveal significant RV variability. 
The measured RVs are listed in Tables\,\ref{tab:bds} and \ref{tab:tts}
and plotted in Figs.\,\ref{fig:rvs1}--\ref{fig:rvs5}. They are presented in 
detail in Sect.\,\ref{sect:rvconst} and \ref{sect:rvvari} after
elaborating on the probed companion masses and orbital separations in the following section.

\subsection{Covered parameter space}

\begin{figure}[t]
\centering
\includegraphics[width=\linewidth]{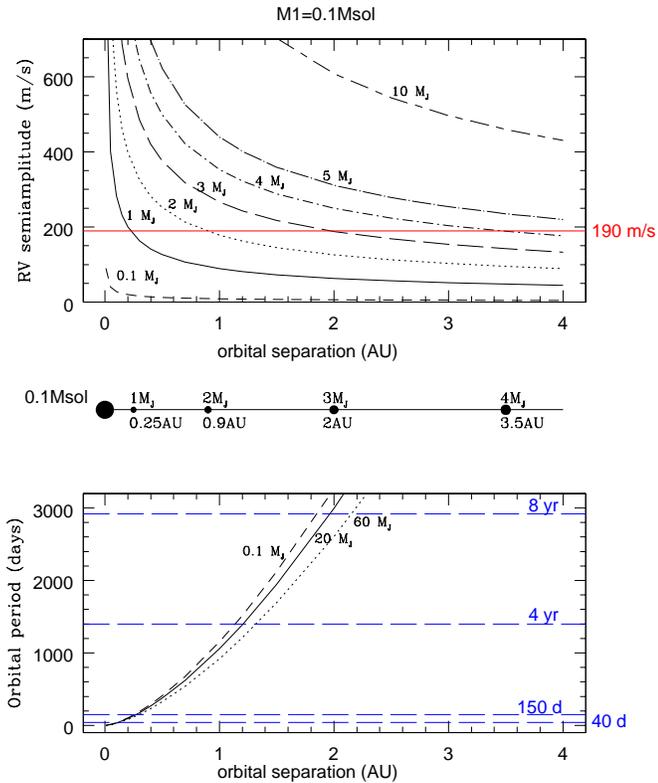}
\caption{
\label{fig:sep}
Covered orbital separation ranges as function of RV precision
and time base of the obtained data for a 0.1\,M$_{\odot}$ primary.
\emph{Top panel:} 
RV semiamplitude vs orbital separation for different companion masses.
With the average RV precision of 190\,m\,s$^{-1}$ for primaries of 
0.1\,M$_{\odot}$ achieved in this survey,
a 1\,M$_\mathrm{Jup}$ companion can be detected out to 0.25\,AU,
a 2\,M$_\mathrm{Jup}$ companion out to 0.9\,AU, a 3\,M$_\mathrm{Jup}$ companion
out to 2\,AU, a 4\,M$_\mathrm{Jup}$ companion out to 3.5\,AU, and 
companions of $\geq$ 5\,M$_\mathrm{Jup}$ at least out to 4\,AU.
However, due to a limited time base not all of the possible separation ranges
have been covered yet. This is displayed in the 
\emph{bottom panel}, which shows the orbital period vs separation
exemplarily for companion masses of 0.1\,M$_\mathrm{Jup}$,
20\,M$_\mathrm{Jup}$, and 60\,M$_\mathrm{Jup}$.  
The orbital periods sampled in this survey so far 
vary among the different targets. For all targets, a period of 
40 days has been covered,
allowing the detection of substellar companions out to
0.1\,AU. For some targets, periods of 150 days, 
4\,yr, and 8\,yr have also been probed
and, thus, correspondingly larger separations, as indicated in the plot.
}
\end{figure}
\begin{figure}[t]
\centering
\includegraphics[width=\linewidth]{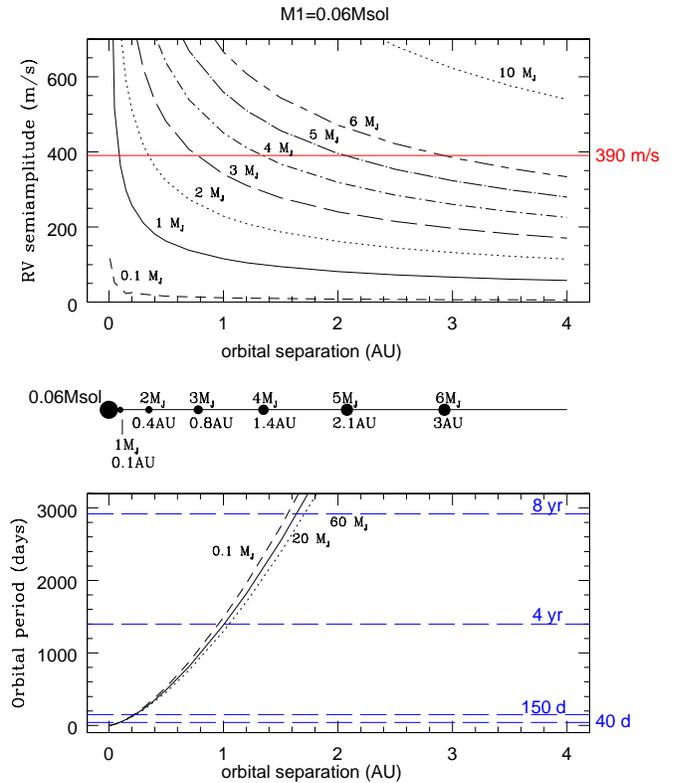}
\caption{
\label{fig:sep60Mj}
Same as Fig.\,\ref{fig:sep} but for a 
0.06\,M$_{\odot}$ primary for which the 
RV precision achieved in this survey is 390\,m\,s$^{-1}$ on average.
}
\end{figure}

For Cha\,H$\alpha$\,1, 2, 3, 5, 6, 7, 8, and 12,
two RV measurements were obtained with a $\sim$20 day time offset, 
while for Cha\,H$\alpha$\,4, Cha\,H$\alpha$\,8, B\,34, CHXR\,74, and Sz\,23, 
four to five RV points were obtained with sampling intervals between
5 and 70 days.
In addition, follow-up observations were also performed for four targets:
for Cha\,H$\alpha$\,4, Cha\,H$\alpha$\,8 after two years and 
for CHXR\,74, Sz\,23 after four years.

Figures \ref{fig:sep} and \ref{fig:sep60Mj} illustrate the probed parameter space in terms of 
companion masses, orbital separations, and periods for primaries of 
0.1\,M$_{\odot}$ and 0.06\,M$_{\odot}$, respectively.
For targets with a mass of about 0.1\,M$_{\odot}$ (Cha\,H$\alpha$\,4, 5, B\,34),
an \emph{average} RV precision of 190\,m\,s$^{-1}$ was achieved in this survey.
It can be seen from Fig.\,\ref{fig:sep} (upper panel and middle sketch)
that this basically allows a 1\,M$_\mathrm{Jup}$ companion
to be detected at separations 
of 0.25\,AU or smaller, while a 3\,M$_\mathrm{Jup}$ companion can be detected 
out to 2\,AU, and companions with $\geq$5\,M$_\mathrm{Jup}$ at least out to 
4\,AU\footnote{It is noted that the average RV precision for B\,34 alone is 
significantly better (90\,m\,s$^{-1}$) and, therefore, allows detection of less massive 
companions at larger separations, e.g. 1\,M$_\mathrm{Jup}$ out to 1\,AU.}.
However, due to a limited time base, the whole   
possible separation range has not been covered yet.
The orbital periods sampled in this survey so far 
vary among the different targets and 
are between 40 days (a time interval of 20 days allows detection of an orbital
period twice as long) and 8 years. As displayed in the lower panel of Fig.\,\ref{fig:sep},
this period sampling allows substellar companions to be detected 
at separations between 0.1\,AU and about 2\,AU relatively independent of companion mass.

Figure \ref{fig:sep60Mj} shows the situation for a lower mass primary, here
for 0.06\,M$_{\odot}$.
With the average RV precision of 390\,m\,s$^{-1}$ achieved for primaries of about 
(0.06$\pm0.01$)\,M$_{\odot}$
(Cha\,H$\alpha$\,2, 3, 6, 8, 12), a 1\,M$_\mathrm{Jup}$ companion is basically detectable 
out to 0.1\,AU, a 3\,M$_\mathrm{Jup}$ companion out to 0.8\,AU and companions with
$\geq$6\,M$_\mathrm{Jup}$ at least out to 3\,AU.
Based on the sampled orbital periods between 40 days and 8 years,
companions could have been detected so far out to 0.1\,AU for all targets in this mass
range and out to about 1.6\,AU for Cha\,H$\alpha$\,8
(Fig.\,\ref{fig:sep60Mj}, lower panel).
While the RV signal caused by an orbiting companion
of a certain mass is generally larger for a lower mass primary than for a higher mass one,
the significantly lower average RV precision
for primaries of $\sim$0.06\,M$_{\odot}$ (390\,m\,s$^{-1}$) compared to 0.1\,M$_{\odot}$ (190\,m\,s$^{-1}$)
restricts the accessible separation range. 

To summarize, for all targets, separations $\la$0.1\,AU ($\sim$40 day period) have been probed.
In addition, for Cha\,H$\alpha$\,4 and B\,34 separations $\la$0.25\,AU ($\sim$150 days) have been
studied, 
for Cha\,H$\alpha$\,4 and Cha\,H$\alpha$\,8 $\la$1--1.2\,AU (4 yr)
and for CHXR\,74 and Sz\,23 $\la$2.5\,AU (8\,yr).

\subsection{RV constant objects}
\label{sect:rvconst}

\begin{figure}[h]
\begin{center}
\includegraphics[width=.45\textwidth]{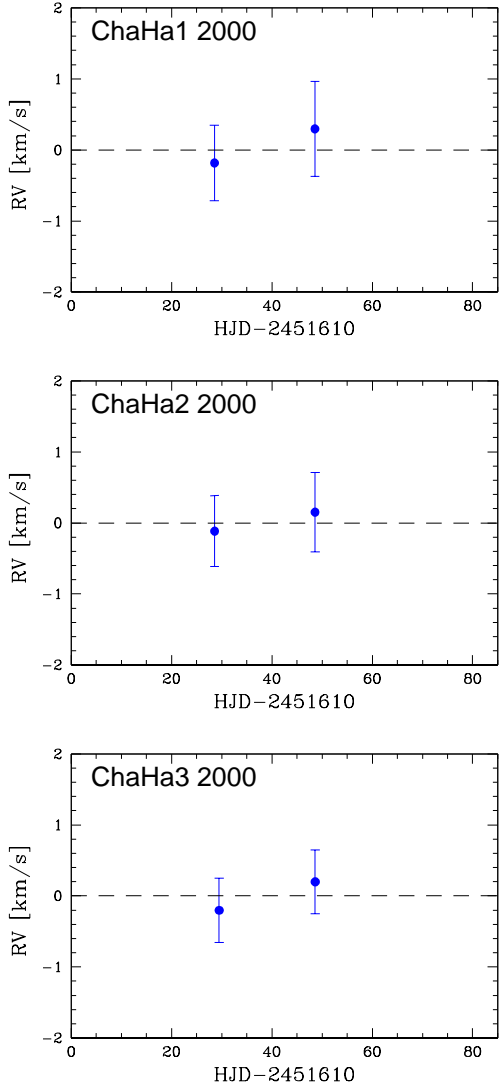}
\end{center}
\caption{
\label{fig:rvs1} 
RV constant objects: relative RV vs. time in Julian days for BDs/VLMSs in Cha\,I
based on high-resolution UVES/VLT spectra. Error bars indicate 1 $\sigma$ errors.
}
\end{figure}

\begin{figure}[t]
\begin{center}
\includegraphics[width=.45\textwidth]{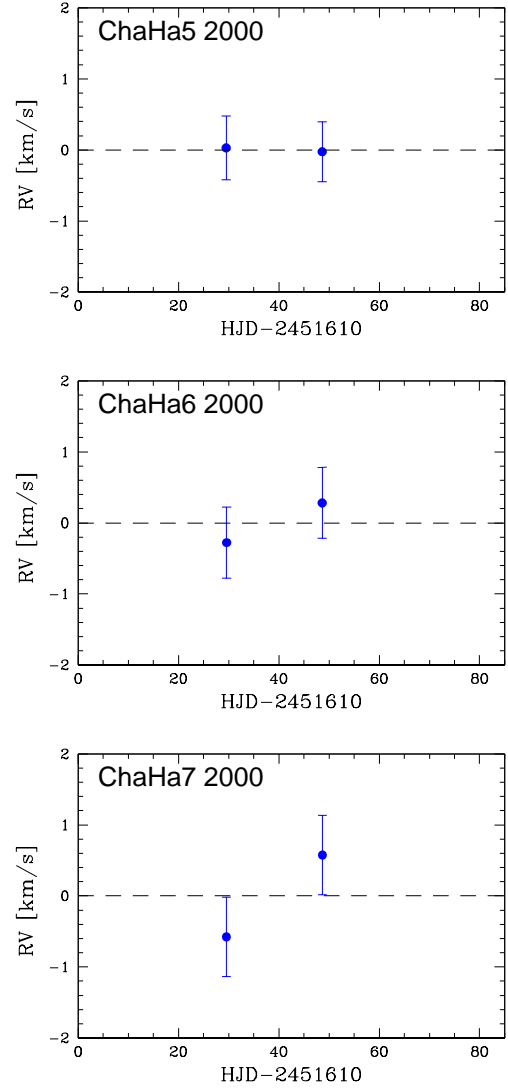}
\end{center}
\caption{
\label{fig:rvs2} 
RV constant objects continued. See Fig.\,\ref{fig:rvs1}.
}
\end{figure}

\begin{figure}[t]
\begin{center}
\includegraphics[width=.45\textwidth]{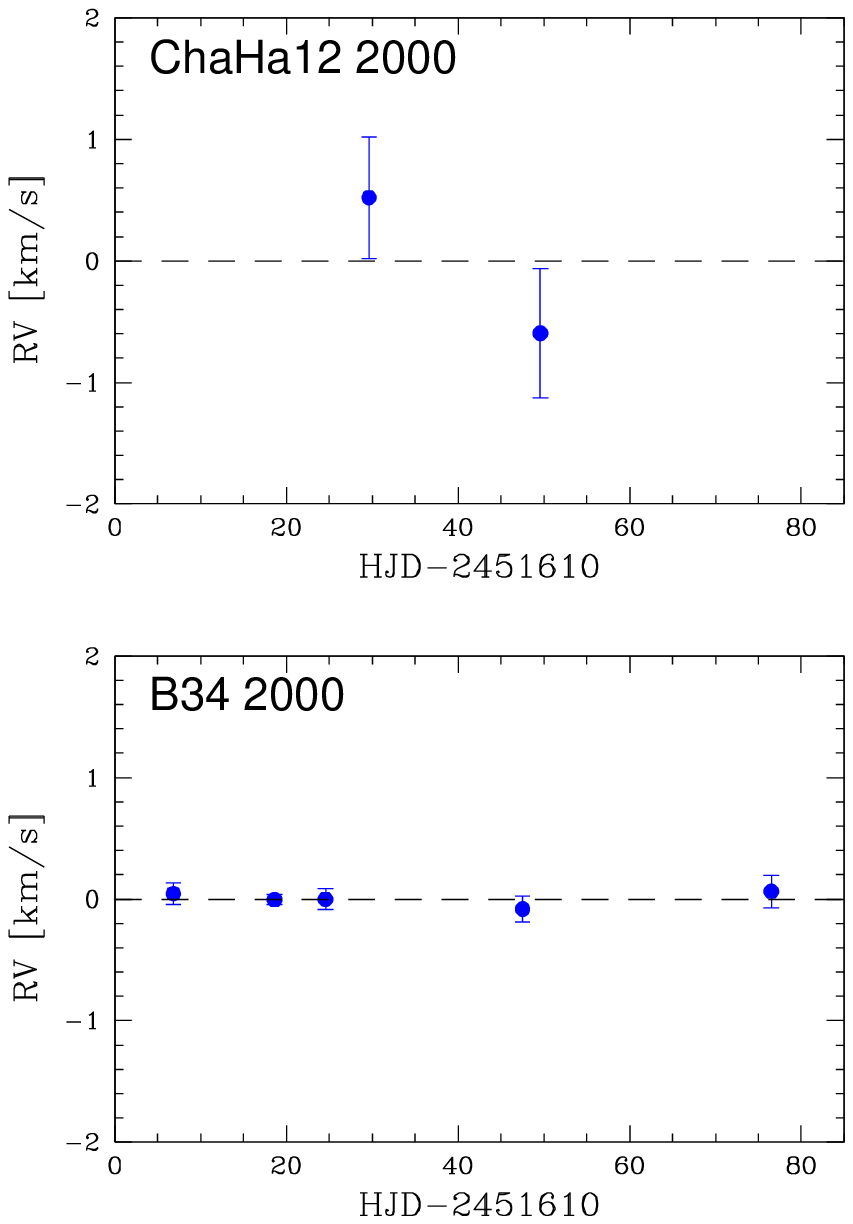}
\end{center}
\caption{
\label{fig:rvs3} 
RV constant objects continued. See Fig.\,\ref{fig:rvs1}.
}
\end{figure}

\begin{figure*}[t]
\begin{center}
\includegraphics[width=.9\textwidth]{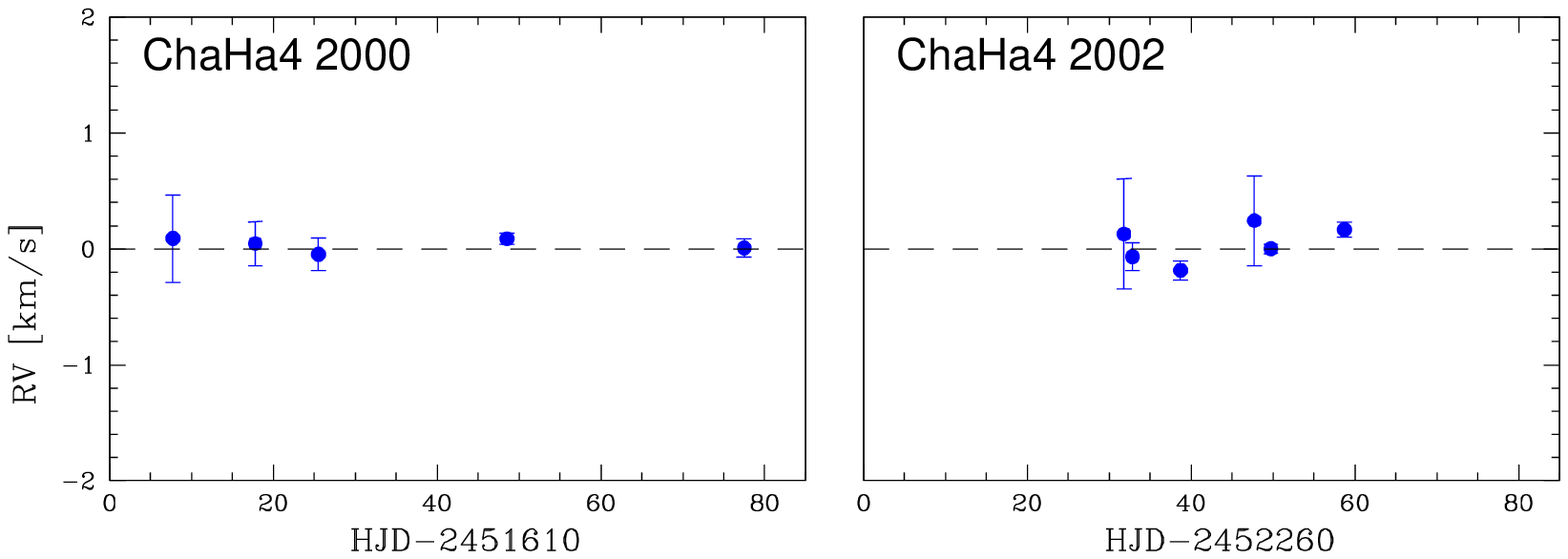}
\end{center}
\caption{
\label{fig:rvs4} 
RV constant objects continued. See Fig.\,\ref{fig:rvs1}.
}
\end{figure*}

The RVs for the BDs/VLMSs
Cha\,H$\alpha$\,1, 2, 3, 4, 5, 6, 7, 12, and B\,34 are constant within the 
measurement uncertainties of 2\,$\sigma$ for Cha\,H$\alpha$\,4 and of 1\,$\sigma$ for 
all others (Figs.\,\ref{fig:rvs1}--\ref{fig:rvs4}).
The sampling intervals vary among these objects. For the majority, it 
is $\la$20 days corresponding to $\la$0.1\,AU, while it is $\la$70 days ($\la$0.25\,AU)
for B\,34  and $\la$ two years ($\la$1.2\,AU) for Cha\,H$\alpha$\,4.
From the non-detections of variability, we estimated upper limits for 
the projected masses M$_2 \sin i$ of hypothetical companions for each object.
Table\,\ref{tab:upperlimits} lists RV differences
for all monitored targets.
The upper limits for M$_2 \sin i$ of hypothetical companions around the
RV constant BDs/VLMSs range between 0.1\,M$_\mathrm{Jup}$ and 1.5\,M$_\mathrm{Jup}$ 
(Table\,\ref{tab:upperlimits}, upper part)
assuming a 
circular orbit, a separation of 0.1\,AU between companion and primary object, and
adopting primary masses from Comer\'on et al. (1999, 2000).
The adopted orbital separation of 0.1\,AU corresponds to
orbital periods ranging between 30 and 70 days for them

As discussed in Guenther \& Wuchterl (2003),
the snow-radius (i.e. the smallest orbital separation at which dust in a surrounding 
disk can condensate
and giant planet formation by the core accretion model can occur)
corresponds to orbital periods of 20--40 days for BDs/VLMSs 
as primaries.
Thus, the 0.1\,AU separation adopted by us corresponds to about the 
snow-radius but is sometimes larger.
This means that these nine BDs/VLMSs 
with spectral types M5--M8 and mass estimates $\la$0.12\,M$_{\odot}$
show no RV variability down to Jupiter mass planets for separations $\la$0.1\,AU
(0.25\,AU for B\,34 and 1.2\,AU for Cha\,H$\alpha$\,4).
There is, of course,
the possibility that existing companions have not been detected
due to non-observations at the critical orbital phases.
Furthermore, long-period companions may have been missed, since 
the sampled orbital periods for all of them but Cha\,H$\alpha$\,4 do not
exceed 5 months.


\subsection{RV variable objects}
\label{sect:rvvari}

\begin{figure*}[t]
\begin{center}
\includegraphics[width=.9\textwidth]{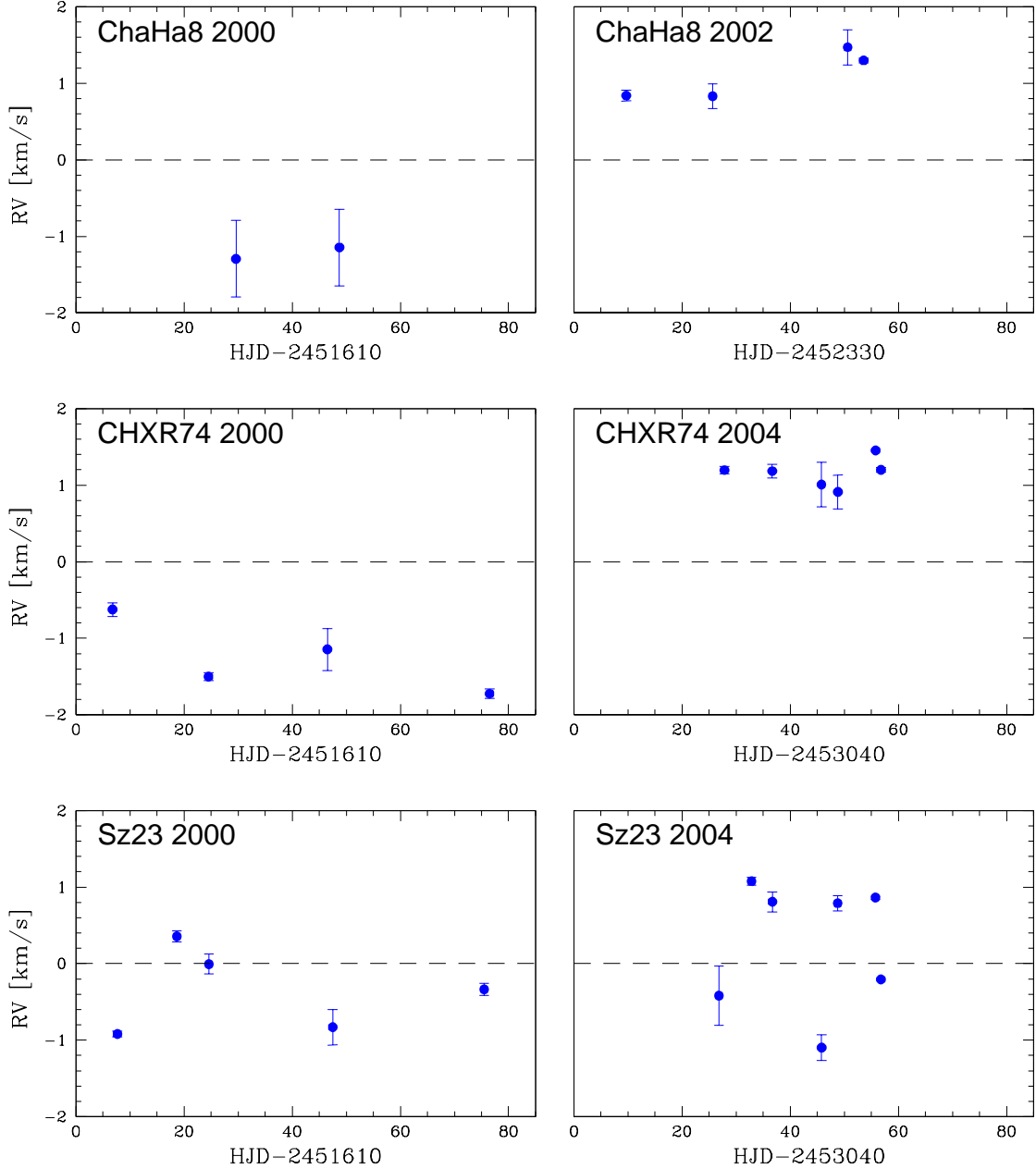}
\end{center}
\caption{
\label{fig:rvs5} 
RV variable objects: 
relative RV vs. time in Julian days for the BD candidate Cha\,H$\alpha$\,8
and the low-mass stars CHXR\,74 and Sz\,23 in Cha\,I
based on UVES/VLT spectra. Error bars indicate 1 $\sigma$ errors.
}
\end{figure*}

For three objects, we found significant RV variations, namely for the BD 
candidate Cha\,H$\alpha$\,8 (M6.5) and for the low-mass stars CHXR\,74 (M4.5) and Sz\,23 (M2.5),
as shown in Fig.\,\ref{fig:rvs5}.
The cross-correlation function for all of them is single-peaked.
The variability characteristic differs among the three objects.
Sz\,23 shows variability on time scales of days with no difference 
in the mean values of RVs recorded in 2000 and in 2004. 
On the other hand, Cha\,H$\alpha$\,8 and CHXR\,74
show no or only smaller amplitude variations
on time scales of days to weeks,
whereas the mean RV measured in 2000 differs significantly 
from the one measured years later, namely in 2002 for Cha\,H$\alpha$\,8 and 
in 2004 for CHXR\,74, respectively, hinting at variability periods on
the order of months or longer.  

One possible explanation of the nature of these RV variations is that they 
are caused by surface activity, since prominent surface spots
can cause a shifting of the photo center at the rotation period.
The upper limits for the rotational periods of Cha\,H$\alpha$\,8,
CHXR\,74, and Sz\,23 are 1.9\,d, 4.9\,d, and 2.1\,d, 
based on projected rotational velocities $v \sin i$ (Joergens \& Guenther 2001, 
cf. also Joergens et al. 2003).
Thus, the time-scale of the RV variability of Sz\,23 is on the order 
of the rotation period and could be a rotation-induced phenomena, while
the RV variability of Cha\,H$\alpha$\,8 and CHXR\,74 on time scales of months to years
cannot be explained as rotational modulation. 

The other possibility is that the RV variations are the 
Doppler shift caused by the gravitational force of orbiting companions.
The poor sampling does not allow us to determine periods of the variations,
but we can make some useful estimates.
Based on the data for Cha\,H$\alpha$\,8 ($\sim$0.07\,M$_{\odot}$), 
we suggest that its RV period is at least 
150 days, which transfers to an orbital separation of $\ga$ 0.2\,AU,
only weakly depending on the companion mass (cf. Fig.\,\ref{fig:sep60Mj}).
The recorded half peak-to-peak RV difference of 1.4\,km\,s$^{-1}$
is a lower limit for the RV semiamplitude caused by a hypothetical companion.
Thus, a companion causing these variations has to have a mass M$_2 \sin i$ of at least 
6\,M$_\mathrm{Jup}$ when assuming a circular orbit.
For CHXR\,74, a period of $\ga$ 200 days would be consistent with the 
RV data of 2000 and 2004 and would correspond to 
a separation of $\ga$ 0.4\,AU and, thus, to
a companion with mass $\ga$15\,M$_\mathrm{Jup}$.

%


\subsection{RV noise}

\begin{figure}[t]
\centering
\includegraphics[width=\linewidth]{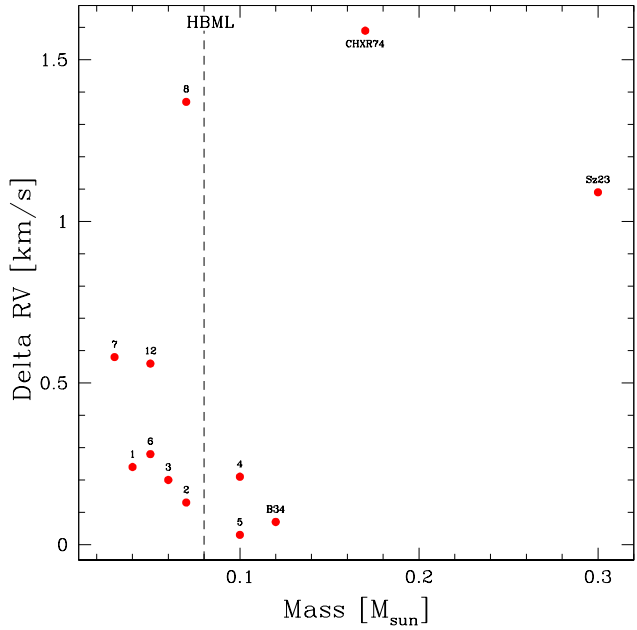}
\caption{
\label{fig:rvmass}
RV variations vs. object mass.
Plotted are half peak-to-peak differences of the observed RVs.
Each data point is labeled with the corresponding object name, the numbers
denote the Cha\,H$\alpha$ objects.
The upper three data points represent the RV variable objects with $\Delta$RV above 
1\,km\,s$^{-1}$.
The remaining data points represent RV constant objects.
The decrease of $\Delta$RV with increasing mass for the latter group
indicates that they display no
significant RV noise due to activity, 
which would cause the opposite effect, i.e.
an increasing RV amplitude with mass.
Mass estimates are from Comer\'on et al. (1999, 2000).
HBML roughly indicates the theoretical 'hydrogen burning mass limit'.
}
\end{figure}
\begin{figure}[t]
\centering
\includegraphics[width=\linewidth]{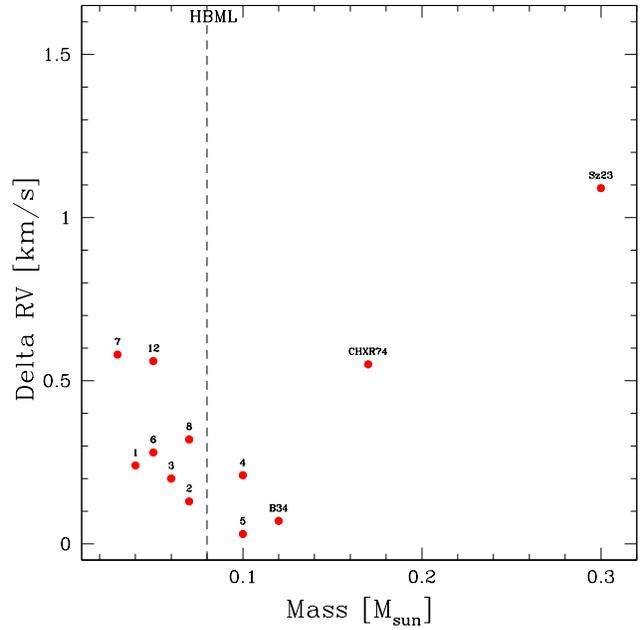}
\caption{
\label{fig:rvmass_short}
RV noise level: Same as Fig.\,\ref{fig:rvmass} but only variations
on time scales of days to weeks are considered.
Differences to Fig.\,\ref{fig:rvmass} occur   
for Cha\,H$\alpha$\,8 and CHXR\,74, whereas for all other objects, either data have been
recorded anyway only within one year or the RV variations observed on
the short- and long-term do not differ in the recorded amplitude. 
It can be seen that the downward trend is reversed somewhere between
0.1 and 0.2\,M$_{\odot}$.
}
\end{figure}

Figure \ref{fig:rvmass} displays the measured 
difference in velocity $\Delta$RV for all targets of the RV survey
in terms of the half size of the peak-to-peak
difference versus their mass,
as adopted from Comer\'on et al. (1999, 2000). 
For the three RV variable objects, $\Delta$RV is above 1\,km\,s$^{-1}$ 
(top three data points 
in Fig.\,\ref{fig:rvmass}), and they are clearly separated from the RV constant objects
in this diagram.
The apparent decrease in $\Delta$RV with increasing mass observed for the 
RV constant objects reflects the dependence of the RV precision on the 
S/N of the spectra. 
Interestingly, it also shows that this group of BDs/VLMSs 
with masses of 0.12\,M$_{\odot}$ and below display no
significant RV noise due to activity, which would cause
systematic RV errors with a RV amplitude increasing with mass.
Two of the variable objects (Cha\,H$\alpha$\,8 and CHXR\,74) only 
show RV variations above 1\,km\,s$^{-1}$ on longer time scales,
while their short-term RV differences are smaller. 
Therefore, we plot $\Delta$RV recorded within one year 
(more precisely with time bases between 20 and 70 days)
for each object in Fig.\,\ref{fig:rvmass_short}.
For Cha\,H$\alpha$\,1, 2, 3, 5, 6, 7, 12, B34, there is no difference
to Fig.\,\ref{fig:rvmass},
since data for these objects were generally recorded only within one year.
For Cha\,H$\alpha$\,4 and Sz\,23, there is also no difference because the 
recorded amplitudes of the short-term and long-term variations
do not differ. However, a difference occurs for Cha\,H$\alpha$\,8 and CHXR\,74 
because of the different $\Delta$RV on the different time scales for these objects.  
Figure \ref{fig:rvmass_short} shows that the relation of decreasing RV difference
with increasing mass in the BD regime 
continues to about 0.12\,M$_{\odot}$ and that it is reversed for higher masses. 
We therefore conclude that BDs/VLMSs
in Cha\,I display no significant RV noise in the mass range 
below about 0.1\,M$_{\odot}$ and that 
roughly between 0.1 and 0.2\,M$_{\odot}$ the activity induced RV noise is increasing
drastically.

\subsection{Multiplicity}

The present RV data hint at a very small multiplicity fraction of very young 
BDs/VLMSs 
for orbital periods roughly $\la$40 days and separations $\la$0.1\,AU.
All ten BDs/VLMSs ($\la$0.12\,M$_{\odot}$) 
in this survey, are RV constant with respect to companions in this parameter range
in our observations.
Among this subsample, there is 
only one (Cha\,H$\alpha$\,8) that shows signs of RV variability, namely on
time scales of at least several months, corresponding to a separation of 
0.2\,AU or larger. This separation range was probed so far for only
two BDs/VLMSs,
so no estimates of 
multiplicity rates in this separation range can be given yet.

The low-mass star CHXR\,74 ($\sim$0.17\,M$_{\odot}$)
shows similar variability behavior as Cha\,H$\alpha$\,8, i.e. 
small amplitude 
variations or no variations on time scales of days/weeks
and larger amplitude RV variability only on longer time scales of at least 
several months.
From Figs.\,\ref{fig:rvmass} and
\ref{fig:rvmass_short},
it can be seen that the recorded long-term RV amplitudes of Cha\,H$\alpha$\,8 and 
CHXR\,74 are significantly above the RV noise level observed on short time scales.
Furthermore, the timescales of the variability are  
much too long to be caused by rotational modulation since the 
rotational periods are on the order of 2 days.
The only other explanation could be a companion with a mass of several Jupiter masses
or more, i.e. a supergiant planet or a brown dwarf.
These observations hint that companions to young BDs and (very) low-mass stars
might have periods of several months or longer, i.e. reside at orbital 
separations outside the snow-radius (cf. Guenther \& Wuchterl 2003).

The RV survey probes the regions close to the central objects, in respect to the occurrence 
of companions, for all targets out to 0.1\,AU and for some even further out 
(2\,AU maximum). 
At the moment, the limits in the separation range covered is set by the time base
rather than by the RV precision. Therefore, follow-up RV measurements are planned.
The small multiplicity fraction found for BDs/VLMSs 
in Cha\,I at small separations ($\la$0.1\,AU) in this RV survey
is also supported by the results of a direct imaging search for 
wide\footnote{The exact separation ranges covered depend on the mass and are e.g.
$>$50\,AU for a 20\,M$_\mathrm{Jup}$ in orbit around a 60\,M$_\mathrm{Jup}$,
or $>$300\,AU for a 1\,M$_\mathrm{Jup}$ in orbit around a 60\,M$_\mathrm{Jup}$.}
(planetary or brown dwarf) companions to mostly the same targets, namely Cha\,H$\alpha$\,1--12, 
by Neuh\"auser et al. (2002, 2003), who find a multiplicity fraction of $\la$10\%.
  
The absolute RVs that are determined based on UVES spectra for Cha\,H$\alpha$\,1, 2, 3, 4, 5, 7, Sz\,23, B\,34, and
CHXR\,74 are consistent with moderately precise RVs measured by Neuh\"auser \& Comer\'on (1999) 
based on medium-resolution spectra within 1.2 times the errors.
The RV measured by these authors for Cha\,H$\alpha$\,8, for which we find 
significant RV variability, is also discrepant 
by 1.6 times the errors with the RV we derived for this object in 2004.
Furthermore, the RV they find for Cha\,H$\alpha$\,6 is discrepant with our value
by 1.9 times the errors, which might be a hint of a spectroscopic companion also 
around Cha\,H$\alpha$\,6.  

\subsection{Brown dwarf desert}

It was one of the surprising results of the RV surveys for extrasolar planets 
around solar-like stars that there is an almost complete absence of companions with 
BD masses in close orbits
($<$3--5\,AU), while at least 6.6\% of the solar-like stars have planets ($<$13\,M$_\mathrm{Jup}$)
and about 15\% of them have stellar companions 
(e.g. Halbwachs et al. 2000, Mazeh et al. 2003, Marcy et al. 2005b).
Whatever physical reason causes the BD desert in the formation of solar-like stars, 
it should also be present in terms of mass ratio 
for primaries of significantly lower or higher mass, as long as they
are formed by the same mechanism.  
For example, if the formation of BDs is a scaled-down low-mass star formation process
(i.e. cloud fragmentation and direct collapse of small cloud cores above the opacity limit),
the BD desert should be found in a scaled-down version shifted to lower companion masses
also around them.

In order to quantify this, we looked at the distribution of mass ratios for known RV planets 
around solar-like stars and inferred from it that the lower boundary 
of the BD desert is at M$_2$/M$_1\approx$0.02.
Its upper boundary, on the other hand, is not as well-defined. 
The lowest mass ratio
known for a \emph{stellar} spectroscopic companion to a solar-like star is 0.2 (Prato et al. 2002,
Mazeh et al. 2003); however,
as pointed out by Mazeh et al. (2003),
stellar systems with M$_2$/M$_1<$0.2-0.3 have not been studied well yet.
For the following consideration, 
a mass ratio of 0.08 is
somewhat arbitrarily assumed as the upper value for the BD desert,
but this is not confirmed by observations.

The BD desert around solar-like stars 
is now be scaled down to the primary masses of 
the target BDs and VLMSs of this work
(0.03\,$\la\,\mbox{M}_1\,\la$\,0.12\,M$_{\odot}$).
For the lowest mass primary studied here (0.03\,M$_{\odot}$), the BD desert would 
be shifted towards companion masses of 0.6--2.5\,M$_\mathrm{Jup}$, and for
a 0.12\,M$_{\odot}$ primary towards 2.5--10\,M$_\mathrm{Jup}$.
Thus, the scaled-down equivalent to the BD desert around solar-like stars would be 
a \emph{giant planet desert} around BD and VLMS primaries. 

Our RV survey started to test its presence 
around the targets in Cha\,I.
For the orbital separations covered so far (for all targets $\la$0.1\,AU and 
for some even larger with a maximum of $\la$1.2\,AU for M$_1\la$0.12\,M$_{\odot}$),  
companions in the `giant planet desert' are clearly detectable for primaries above 
0.06\,M$_{\odot}$, 
whereas for the lowest mass primaries studied (0.03--0.05\,M$_{\odot}$) the 
sensitivity allows only a part of the `giant planet desert' around them to be probed.
No hints of companions within these parameter ranges were found in the 
data. The lower mass limits roughly estimated for hypothetical companions
around the RV variable objects Cha\,H$\alpha$\,8 and CHXR\,74 would locate them just  
outside the `giant planet desert' around them, which is 1.5--5.9\,M$_\mathrm{Jup}$
for Cha\,H$\alpha$\,8 and 4--14\,M$_\mathrm{Jup}$ for CHXR\,74, respectively;
however, the assumed upper boundary is somewhat uncertain, as pointed out above.

So far, only a fraction of the orbital separations,
for which the BD desert is established around solar-like stars ($<$3--5\,AU),
has been probed yet.
Larger separations will be explored by follow-up UVES observations.

For higher than solar-mass primaries, 
RV surveys of K giants detected a much higher rate of 
close BD companions compared to solar-like stars 
(Frink et al. 2002, Hatzes et al. 2005, Setiawan 2005, Mitchell et al. 2005), 
which with only one exception
all correspond to mass ratios $<$0.02, i.e. do not lie in the brown dwarf
desert when scaled up for the higher primary masses.

\begin{table*}[t]
\caption{
\label{tab:bds} 
RVs for BDs and VLMSs (M6--M8) in Cha\,I.
Given are the date of the observation,
heliocentric Julian day (HJD) at the middle of the exposure,
the measured RV and the estimated error $\sigma_{RV}$ of the relative RVs.
The asterisk marks RVs based on the average of two single measurements for which 
the errors are standard deviations.
The last column gives the weighted mean RV for the individual objects and the 
error of this mean, which takes into account an uncertainty of 400\,m\,s$^{-1}$ for the
zero point of the velocity.  
}
\vspace{0.1cm}
\begin{tabular}{lcccl|l}
\hline
\hline
\myrule
Object             &   Date   & HJD & RV   & ~$\sigma_{RV}$          & $\overline\mathrm{RV}$ \\
                   & & & [km\,s$^{-1}$] & [km\,s$^{-1}$]             & [km\,s$^{-1}$] \\
\hline
\hline
Cha\,H$\alpha$\,1    & 2000 Apr 04 & 2451638.56395 & 16.167 & ~0.53 \\ 
		     & 2000 Apr 24 & 2451658.57346 & 16.648 & ~0.67 & 16.35 $\pm$ 0.63 \\
\hline
Cha\,H$\alpha$\,2    & 2000 Apr 04 & 2451638.59431 & 16.015 & ~0.50 \\
		     & 2000 Apr 24 & 2451658.60407 & 16.282 & ~0.56 & 16.13 $\pm$ 0.53 \\
\hline
Cha\,H$\alpha$\,3    & 2000 Apr 04 & 2451639.49340 & 14.357 & ~0.45 \\
		     & 2000 Apr 24 & 2451658.61991 & 14.758 & ~0.45 & 14.56$\pm$ 0.60 \\
\hline 
Cha\,H$\alpha$\,4    
		   & 2000 Mar 14   & 2451617.73646 & 14.909 & ~0.38\,* \\

		   & 2000 Mar 24   & 2451627.80388 & 14.866 & ~0.19\,* \\

		   & 2000 Mar 31   & 2451635.51085 & 14.773 & ~0.14 \\ 

	           & 2000 Apr 23   & 2451658.52150 & 14.908 & ~0.05\,* \\

	           & 2000 May 22   & 2451687.50595 & 14.830 & ~0.08\,* & 14.82 $\pm$ 0.40\\

	    	   & 2002 Jan 17   & 2452291.78912 & 14.949 & ~0.48\,* \\

		   & 2002 Jan 18   & 2452292.82508 & 14.754 & ~0.12\,* \\

         	   & 2002 Jan 24   & 2452298.70540 & 14.635 & ~0.08\,* \\

		   & 2002 Feb 02   & 2452307.71979 & 15.064 & ~0.39\,* \\ 

		   & 2002 Feb 04   & 2452309.74222 & 14.821 & ~0.001\,* \\

		   & 2002 Feb 13   & 2452318.71915 & 14.985 & ~0.07\,* \\
\hline
Cha\,H$\alpha$\,5    & 2000 Apr 05 & 2451639.51485 & 15.499 & ~0.45 \\
		     & 2000 Apr 24 & 2451658.63522 & 15.446 & ~0.42 & 15.47$\pm$ 0.43\\
\hline
Cha\,H$\alpha$\,6    & 2000 Apr 05 & 2451639.58967 & 16.093 & ~0.50 \\
		     & 2000 Apr 24 & 2451658.65099 & 16.652 & ~0.50 & 16.37$\pm$ 0.68\\
\hline
Cha\,H$\alpha$\,7    & 2000 Apr 05 & 2451639.55225 & 16.513 & ~0.56 \\
		     & 2000 Apr 24 & 2451658.68756 & 17.664 & ~0.56 & 17.09$\pm$ 0.98\\
\hline
Cha\,H$\alpha$\,8    & 2000 Apr 05 & 2451639.61095 & 14.787 & ~0.50 & \\
		     & 2000 Apr 24 & 2451658.72597 & 14.935 & ~0.50 & 14.86 $\pm$ 0.47 (2000)\\
                     & 2002 Mar 06 & 2452339.68965 & 16.920 & ~0.07\,* \\ 
                     & 2002 Mar 22 & 2452355.65264 & 16.912 & ~0.16\,* \\
                     & 2002 Apr 16 & 2452380.61646 & 17.551 & ~0.23\,* & \\              
                     & 2002 Apr 19 & 2452383.57565 & 17.379 & ~0.03\,* & 17.30 $\pm$ 0.50 (2002)\\
\hline
Cha\,H$\alpha$\,12   & 2000 Apr 05 & 2451639.63487 & 15.021 & ~0.50 \\
		     & 2000 Apr 25 & 2451659.59469 & 13.905 & ~0.53 & 14.50$\pm$ 0.96\\
\hline
\hline
\end{tabular}
\end{table*}


\begin{table*}[t]
\caption{
\label{tab:tts}
RVs for (very) low-mass stars (M2.5--M5) in Cha\,I.
See Table\,\ref{tab:bds} for details.
}
\vspace{0.1cm}
\begin{tabular}{lcccl|l}
\hline
\hline
\myrule
Object              &  Date        & HJD         & RV   & ~$\sigma_{RV}$  & $\overline\mathrm{RV}$ \\
                    &              &             & [km\,s$^{-1}$] & [km\,s$^{-1}$] & [km\,s$^{-1}$]\\
\hline
\hline
\myrule
B\,34                & 2000 Mar 13 & 2451616.83205& 15.795 & ~0.09\,*\\

		     & 2000 Mar 25 & 2451628.61377& 15.746 & ~0.04 \\

		     & 2000 Mar 31 & 2451634.55482& 15.749 & ~0.09\,* & 15.75 $\pm$ 0.42 \\

		     & 2000 Apr 23 & 2451657.53470& 15.667 & ~0.11\,*\\

	    	     & 2000 May 22 & 2451686.51384& 15.814 & ~0.13\,*\\
\hline
CHXR\,74             & 2000 Mar 13 & 2451616.78715 & 15.376 & ~0.09\,* \\
		     & 2000 Mar 31 & 2451634.52092 & 14.499 & ~0.05\,* & \\
		     & 2000 Apr 22 & 2451656.51247 & 14.854 & ~0.27\,* & 14.58 $\pm$ 0.62 (2000)\\
		     & 2000 May 21 & 2451686.48261 & 14.276 & ~0.06\,* \\

                     & 2004 Mar 03 & 2453067.82956 & 17.196 & ~0.05\,* \\
                     & 2004 Mar 12 & 2453076.66488 & 17.184 & ~0.09\,* & \\
                     & 2004 Mar 21 & 2453085.77022 & 17.009 & ~0.29\,* & 17.42 $\pm$ 0.44 (2004)\\
                     & 2004 Mar 24 & 2453088.79822 & 16.912 & ~0.22\,* \\
                     & 2004 Mar 31 & 2453095.78042 & 17.454 & ~0.01\,* \\
                     & 2004 Apr 01 & 2453096.77600 & 17.200 & ~0.03\,* \\
\hline
Sz\,23               & 2000 Mar 14 & 2451617.68093 & 14.652 & ~0.04    \\
		     & 2000 Mar 25 & 2451628.66914 & 15.926 & ~0.07    \\
		     & 2000 Mar 31 & 2451634.59142 & 15.564 & ~0.13\,* \\
		     & 2000 Apr 22 & 2451657.49636 & 14.740 & ~0.23\,* \\
		     & 2000 May 20 & 2451685.48812 & 15.233 & ~0.08\,* \\

                     & 2004 Mar 02 & 2453066.80358 & 15.152 & ~0.39\,* & 15.57 $\pm$ 0.55\\
                     & 2004 Mar 08 & 2453072.84053 & 16.647 & ~0.05\,* \\
                     & 2004 Mar 12 & 2453076.69419 & 16.377 & ~0.13\,* \\
                     & 2004 Mar 21 & 2453085.80059 & 14.472 & ~0.17\,* \\
                     & 2004 Mar 24 & 2453088.77028 & 16.360 & ~0.10\,* \\
                     & 2004 Mar 31 & 2453095.75207 & 16.432 & ~0.02\,* \\
                     & 2004 Apr 01 & 2453096.74897 & 15.364 & ~0.01\,* \\
\hline
\hline
\end{tabular}
\end{table*}


\begin{table*}[t]
\caption{
\label{tab:upperlimits}
RV differences and estimates of companion masses. 
$\Delta$RV gives half peak-to-peak differences of the observed RVs listed in 
Tables\,\ref{tab:bds} and \ref{tab:tts}.
For RV constant objects in this survey (upper part of the table), $\Delta$RV 
is an upper limit for the RV semiamplitude of hypothetical companions missed due to RV precision.
For them, the last column lists upper limits for the companion minimum mass M$_2 \sin i$
derived by assuming a semimajor axis of 0.1\,AU and circular orbits.
For RV variable objects (lower part of table), the recorded RV differences are  
lower limits for the RV amplitudes of hypothetical companions.
For them, the last column gives a lower limit for the companion minimum mass M$_2 \sin i$
for separations of 0.2\,AU (Cha\,H$\alpha$\,8) and 0.4\,AU (CHXR\,74, Sz\,23), resp. 
Primary masses are taken from Comer\'on et al. (1999, 2000). 
}
\vspace{0.1cm}
\begin{tabular}{lcc}
\hline
\hline
\myrule
object& $\Delta$RV  & M$_2 \sin i$ \\
& [km\,s$^{-1}$] & [M$_\mathrm{Jup}$]\\
\hline
\hline
\myrule
RV constant: & & max. M$_2 \sin i$\\
\hline
Cha\,H$\alpha$\,1   & 0.24 & 0.6 \\
Cha\,H$\alpha$\,2   & 0.13 & 0.4 \\
Cha\,H$\alpha$\,3   & 0.20 & 0.6 \\
Cha\,H$\alpha$\,4   & 0.21 & 0.8 \\
Cha\,H$\alpha$\,5   & 0.03 & 0.1 \\
Cha\,H$\alpha$\,6   & 0.28 & 0.6 \\
Cha\,H$\alpha$\,7   & 0.58 & 1.1 \\
Cha\,H$\alpha$\,12  & 0.56 & 1.5 \\
B\,34               & 0.07 & 0.3 \\ 
\hline
RV variable: & & min. M$_2 \sin i$\\
\hline
Cha\,H$\alpha$\,8   & 1.37 & 6 \\
CHXR\,74            & 1.59 & 15 \\
Sz\,23              & 1.09 & 13\\
\hline
\hline
\end{tabular}
\end{table*}


\section{Conclusions and summary}
\label{sect:concl}

We have presented time-resolved high-resolution spectroscopic observations with UVES at the VLT of 
a population of very young BDs and (very) low-mass stars  
in the Cha\,I star-forming region. 
As they have an age of only a few million years, exploration
allows insight into the formation and early evolution of BDs
and of stars close to the substellar borderline.
The RV precision achieved in this RV survey is sufficient for detecting
companions down to Jupiter mass planets. The orbital periods sampled so far 
correspond to orbital separations of $\la$0.1\,AU
(for some also to larger separations up to almost 3\,AU). 
Therefore, it allows us to probe planet formation
at very young ages (1--10\,Myrs), around very low-mass, partly substellar, primaries
and at close orbital separations. 
This combination of primary mass range, primary spectral type, age, and separations
has not been covered by previous surveys, which were either done by direct imaging,
and, therefore, were only sensitive to larger separations or the primaries
were of considerably larger mass or age.

The analysis of the UVES spectra reveals very constant RVs on time scales 
of weeks to months for the majority of the targets, as well as RV variability for 
three sources.
The RV constant objects are six BDs and three VLMSs 
(M$\la$0.12\,M$_{\odot}$, spectral types M5--M8), for which we estimate 
upper limits for masses
of hypothetical companions in the range of 0.1\,M$_\mathrm{Jup}$ to 1.5\,M$_\mathrm{Jup}$
by assuming orbital separations of 0.1\,AU, which corresponds to the sampled periods
(40 days).
The data show that this group displays no significant RV noise due to activity down 
to the precision necessary to detect Jupiter mass planets.
This demonstrates that BDs/VLMSs in Cha\,I 
are suitable targets when using the RV technique to search for planets. 
This opens up a new parameter range for the search for extrasolar planets, namely
very young planets at very close separations. 
Even with the unprecedented angular resolution
of the new generation of optical and near-IR ground-based
interferometers, like the VLT Interferometer, the separation ranges for which the 
RV method is sensitive are not covered by existing and near-future instruments
for planetary mass companions in nearby star forming regions.

Three objects exhibit significant RV variations with peak-to-peak RV differences of 
2--3\,km\,s$^{-1}$, namely
the BD candidate Cha\,H$\alpha$\,8 (M6.5) and the low-mass stars CHXR\,74 
(M4.5, $\sim$0.17\,M$_{\odot}$) and Sz\,23 (M2.5, $\sim$0.3\,M$_{\odot}$).
A possible explanation for the short-term RV variations 
on time scales of days for Sz\,23, which is the highest mass object in the sample, 
are surface spots. 
The other two variable objects, Cha\,H$\alpha$\,8 and CHXR\,74, 
show different variability behavior. They display
only very small or no RV variability on time scales of days to weeks but significant 
RV variations on times scales of months or longer, which cannot be explained by  
rotational modulation and, therefore, it hints at orbiting companions. 
The poor phase coverage does not allow determination of orbital parameters
for the hypothetical companions.
However, the RV data for Cha\,H$\alpha$\,8 suggest that its period is at least five months, 
which would correspond to orbital separations of at least 0.2\,AU. 
Based on these numbers, the detected RV variations of Cha\,H$\alpha$\,8
could be caused by a 6\,M$_\mathrm{Jup}$ or a more massive companion.
For CHXR\,74, the data suggest a period of 7 months or longer ($\ga$0.4\,AU) 
and a companion of at least 15 \,M$_\mathrm{Jup}$.
In order to explore the nature of the detected RV variations, 
follow-up RV measurements of CHXR\,74 and Cha\,H$\alpha$\,8 will be performed.
If confirmed as planetary systems, they would be exceptional, because 
they would contain the lowest mass primaries 
and the first BD with an RV planet. With an age of a few million years,
they would also harbor by far the youngest extrasolar RV planet found to date. This would provide
empirical constraints for planet and BD formation and early evolution.

The RV data presented here indicate that the multiplicity fraction of very young 
BDs and (very) low-mass stars is very small for orbital 
separations below 0.1\,AU, which corresponds to about the snow line around the targets
(Guenther \& Wuchterl 2003).
The subsample of ten BDs/VLMSs with masses 
$\la$ 0.12\,M$_{\odot}$,
are RV-constant for orbital periods below 40 days.
In addition, this is true for Cha\,H$\alpha$\,4 and B\,34 for periods below 150 days and 
for Cha\,H$\alpha$\,4 
for periods below 4 years. Only one object of this group,
namely Cha\,H$\alpha$\,8, turned out to be variable on time scales of at least 150 days.
This object and the higher mass CHXR\,74
(not included in the above considered subsample)
hint at the possibility that companions to young BDs/VLMSs
have periods of at least several months. Such a time scale was not covered 
for a substantial part of the targets. Therefore
a multiplicity rate cannot be determined yet. Follow-up RV measurements will probe
these time scales for the remaining targets.

Furthermore, we show that 
the scaled-down equivalent to the BD desert found around solar-like stars would be 
a `giant planet desert' around BDs/VLMSs if formed
by the same mechanism. 
For example, for a 0.03\,M$_{\odot}$ primary, the deserted companion mass region would be
0.6--2.5\,M$_\mathrm{Jup}$ 
and, for a 0.12\,M$_{\odot}$ primary, 2.5--10\,M$_\mathrm{Jup}$.
The present RV data test the existence of such a `giant planet desert' for the targets.
For the orbital separations covered so far, 
companions in the `giant planet desert' are clearly detectable for primaries above 
0.06\,M$_{\odot}$, 
whereas for the lowest mass primaries studied (0.03--0.05\,M$_{\odot}$), the 
sensitivity allows probing only a part of it.
So far, no hints have been found of companions in these mass ranges.

At much larger separations, a direct imaging search for wide
(planetary or brown dwarf) companions to mostly the same targets 
also found a very small multiplicity fraction (Neuh\"auser et al. 2002, 2003).
There still remains a significant gap in the separation ranges studied, which 
will be probed partly by the planned follow-up RV measurements and is partly only 
accessible with high-resolving AO imaging (NACO\,/\,VLT), or it requires 
interferometric techniques (e.g. AMBER at the VLTI). 

\begin{acknowledgements}
I am grateful to Ralph Neuh\"auser and Eike Guenther
for assistance in the early stages of this project.
I would also like to thank the referee, Fernando Comer\'on, for 
very helpful comments that improved the paper significantly. 
This work made use of software by Eike Guenther to calculate spectroscopic orbits. 
I am pleased to acknowledge the excellent work of the 
ESO staff at Paranal, who took all the UVES observations the present work is based on
in service mode.
I thank Francesca Primas, Ferdinando Patat, Benoit Pirenne, and Eric Louppe 
from ESO Garching for kind and efficient help in various stages of the observation preparation and 
data handling.
Furthermore, I acknowledge a grant by the Deutsche Forschungsgemeinschaft
(Schwerpunktprogramm `Physics of star formation') during the beginning of the project,
as well as current financial support by a Marie Curie Fellowship of the
European Community programme 'Structuring the European Research Area'
under contract number FP6-501875.
Part of the earlier work was carried out at the Max-Planck-Institute for Extraterrestrial Physics,
Garching, Germany.
\end{acknowledgements}

\appendix

\section{Observing logs UVES spectra}

\begin{table}[h]
\caption[Observing log: UVES spectroscopy of BDs and VLMSs in Cha\,I]
{\label{obslog1} 
\small{Observing log: UVES spectroscopy 
of BDs and (very) low-mass stars in Cha\,I.
Given are the date of the observation,
heliocentric Julian day (HJD) at the middle of the exposure,
the exposure time, and the radial velocity RV for each spectrum.
}}
\vspace{0.3cm}
\begin{tabular}{llccc}
\hline
\hline
\myrule
Object             &   Date  & HJD & Exposure & RV             \\
                   &         &     &  [s]     & [km\,s$^{-1}$] \\
\hline
\hline
Cha\,H$\alpha$\,1    & 2000 Apr 04 & 2451638.56395 & 2x1650 & 16.167 \\ 
		     & 2000 Apr 24 & 2451658.57346 & 2x1650 & 16.648 \\
\hline
Cha\,H$\alpha$\,2    & 2000 Apr 04 & 2451638.59431 & 1x1029 & 16.015 \\
		     & 2000 Apr 24 & 2451658.60407 & 1x1029 & 16.282 \\
\hline
Cha\,H$\alpha$\,3    & 2000 Apr 04 & 2451639.49340 & 1x899  & 14.357 \\
		     & 2000 Apr 24 & 2451658.61991 & 1x899  & 14.758 \\
\hline 
Cha\,H$\alpha$\,4    
		   & 2000 Mar 14   & 2451617.72337 & 1x2200 & 15.176 \\
		   & 2000 Mar 14   & 2451617.74954 & 1x2200 & 14.642 \\
		   & 2000 Mar 24   & 2451627.79079 & 1x2200 & 14.732 \\ 
		   & 2000 Mar 24   & 2451627.81697 & 1x2200 & 14.999 \\ 
		   & 2000 Mar 31   & 2451635.51085 & 1x2200 & 14.773 \\ 
		   & 2000 Apr 23   & 2451658.50840 & 1x2200 & 14.941 \\
 		   & 2000 Apr 23   & 2451658.53460 & 1x2200 & 14.875 \\
		   & 2000 May 22   & 2451687.49289 & 1x2200 & 14.774 \\ 
		   & 2000 May 23   & 2451687.51900 & 1x2200 & 14.885 \\ 
		     & 2002 Jan 17 & 2452291.73197 & 2x1100 & 14.761 \\
		     & 2002 Jan 17 & 2452291.75976 & 2x1100 & 15.614 \\
		     & 2002 Jan 17 & 2452291.79010 & 2x1100 & 14.661 \\
		     & 2002 Jan 17 & 2452291.81794 & 2x1100 & 15.260 \\
		     & 2002 Jan 17 & 2452291.84586 & 2x1100 & 14.450 \\
		     & 2002 Jan 18 & 2452292.81113 & 2x1100 & 14.669 \\
		     & 2002 Jan 18 & 2452292.83902 & 2x1100 & 14.839 \\
		     & 2002 Jan 24 & 2452298.69149 & 2x1100 & 14.575 \\
		     & 2002 Jan 24 & 2452298.71931 & 2x1100 & 14.694 \\
		     & 2002 Feb 02 & 2452307.70591 & 2x1100 & 15.338 \\
		     & 2002 Feb 02 & 2452307.73367 & 2x1100 & 14.790 \\
		     & 2002 Feb 04 & 2452309.72831 & 2x1100 & 14.821 \\
		     & 2002 Feb 04 & 2452309.75613 & 2x1100 & 14.820 \\
		     & 2002 Feb 13 & 2452318.70523 & 2x1100 & 14.939 \\
		     & 2002 Feb 13 & 2452318.73306 & 2x1100 & 15.031 \\
\hline
Cha\,H$\alpha$\,5    & 2000 Apr 05 & 2451639.51485 & 1x800  & 15.499 \\
		     & 2000 Apr 24 & 2451658.63522 & 1x800  & 15.446 \\
\hline
Cha\,H$\alpha$\,6    & 2000 Apr 05 & 2451639.58967 & 1x1029 & 16.093 \\
		     & 2000 Apr 24 & 2451658.65099 & 1x1029 & 16.652 \\
\hline
Cha\,H$\alpha$\,7    & 2000 Apr 05 & 2451639.55225 & 2x2150 & 16.513 \\
		     & 2000 Apr 24 & 2451658.68756 & 2x2150 & 17.664 \\
\hline
Cha\,H$\alpha$\,8    & 2000 Apr 05 & 2451639.61095 & 1x1599 & 14.787 \\
		     & 2000 Apr 24 & 2451658.72597 & 1x1600 & 14.935 \\
                     & 2002 Mar 06 & 2452339.67846 & 1x1785 & 16.972 \\
                     & 2002 Mar 06 & 2452339.70084 & 1x1785 & 16.868 \\
                     & 2002 Mar 22 & 2452355.64149 & 1x1785 & 16.795 \\
                     & 2002 Mar 22 & 2452355.66378 & 1x1785 & 17.028 \\
                     & 2002 Apr 16 & 2452380.60528 & 1x1785 & 17.391 \\
                     & 2002 Apr 16 & 2452380.62764 & 1x1785 & 17.710 \\
                     & 2002 Apr 19 & 2452383.56440 & 1x1785 & 17.356 \\
                     & 2002 Apr 19 & 2452383.58690 & 1x1785 & 17.402 \\
\hline
\hline
\end{tabular}
\end{table}

\begin{table}[h]
\caption[Table\,\ref{obslog1} continued]
{\label{obslog2}
\small{Observing log: UVES spectroscopy continued.
See Table\,\ref{obslog1} for details.
}}
\vspace{0.3cm}
\begin{tabular}{llccc}
\hline
\hline
\myrule
Object              &  Date   & HJD & Exposure & RV             \\
                    &         &     &  [s]     & [km\,s$^{-1}$] \\
\hline
\hline
Cha\,H$\alpha$\,12   & 2000 Apr 05 & 2451639.63487 & 1x1599 & 15.021 \\
		     & 2000 Apr 25 & 2451659.59469 & 1x1600 & 13.905 \\
\hline
B\,34                & 2000 Mar 13 & 2451616.82387& 1x1350 & 15.730 \\
		     & 2000 Mar 13 & 2451616.84023& 1x1350 & 15.859 \\
		     & 2000 Mar 25 & 2451628.61377& 1x2700 & 15.746 \\
		     & 2000 Mar 31 & 2451634.54665& 1x1350 & 15.687 \\
		     & 2000 Mar 31 & 2451634.56298& 1x1350 & 15.810 \\
		     & 2000 Apr 23 & 2451657.52652& 1x1350 & 15.591 \\
		     & 2000 Apr 23 & 2451657.54287& 1x1350 & 15.743 \\
		     & 2000 May 21 & 2451686.50566& 1x1350 & 15.720 \\
		     & 2000 May 22 & 2451686.52202& 1x1350 & 15.908 \\
\hline
CHXR\,74             & 2000 Mar 13 & 2451616.78332& 1x600 & 15.313 \\
		     & 2000 Mar 13 & 2451616.79097& 1x600 & 15.439 \\
		     & 2000 Mar 31 & 2451634.51710& 1x600 & 14.537 \\
		     & 2000 Mar 31 & 2451634.52473& 1x600 & 14.460 \\
		     & 2000 Apr 22 & 2451656.50861& 1x600 & 15.048 \\
		     & 2000 Apr 22 & 2451656.51633& 1x600 & 14.660 \\
		     & 2000 May 21 & 2451686.47878& 1x600 & 14.320 \\
		     & 2000 May 21 & 2451686.48644& 1x600 & 14.232 \\

                     & 2004 Mar 03 & 2453067.82538 & 1x630 & 17.234 \\
                     & 2004 Mar 03 & 2453067.83374 & 1x630 & 17.157 \\  

                     & 2004 Mar 12 & 2453076.66087 & 1x630 & 17.123 \\
                     & 2004 Mar 12 & 2453076.66888 & 1x630 & 17.244\\

                     & 2004 Mar 21 & 2453085.76620 & 1x630 & 16.805\\
                     & 2004 Mar 21 & 2453085.77424 & 1x630 & 17.213\\

                     & 2004 Mar 24 & 2453088.79419 & 1x630 & 16.756\\
                     & 2004 Mar 24 & 2453088.80224 & 1x630 & 17.068\\ 

                     & 2004 Mar 31 & 2453095.77639 & 1x630 & 17.447\\ 
                     & 2004 Mar 31 & 2453095.78445 & 1x630 & 17.460\\

                     & 2004 Mar 01 & 2453096.77199 & 1x630 & 17.218\\ 
                     & 2004 Mar 01 & 2453096.78001 & 1x630 & 17.181\\

\hline
Sz\,23               & 2000 Mar 14 & 2451617.68093 & 2x1350 & 14.652 \\
		     & 2000 Mar 25 & 2451628.66914 & 2x1350 & 15.926 \\
		     & 2000 Mar 31 & 2451634.58325& 1x1350 & 15.656  \\
		     & 2000 Mar 31 & 2451634.59958& 1x1350 & 15.472  \\
		     & 2000 Apr 22 & 2451657.48819& 1x1350 & 14.575  \\
		     & 2000 Apr 22 & 2451657.50452& 1x1350 & 14.905  \\
		     & 2000 May 20 & 2451685.47992& 1x1350 & 15.288  \\
		     & 2000 May 20 & 2451685.49632& 1x1350 & 15.178  \\

                     & 2004 Mar 02 & 2453066.79564 & 1x1300 & 15.431\\
                     & 2004 Mar 02 & 2453066.81152 & 1x1300 & 14.873\\

                     & 2004 Mar 08 & 2453072.83264 & 1x1300 & 16.610 \\
                     & 2004 Mar 08 & 2453072.84841 & 1x1300 & 16.684 \\

                     & 2004 Mar 12 & 2453076.68615 & 1x1300 & 16.283 \\
                     & 2004 Mar 12 & 2453076.70222 & 1x1300 & 16.470 \\

                     & 2004 Mar 21 & 2453085.79265 & 1x1300 & 14.591 \\
                     & 2004 Mar 21 & 2453085.80853 & 1x1300 & 14.353 \\

                     & 2004 Mar 24 & 2453088.76236 & 1x1300 & 16.430 \\
                     & 2004 Mar 24 & 2453088.77820 & 1x1300 & 16.289 \\

                     & 2004 Mar 31 & 2453095.74401 & 1x1300 & 16.447 \\
                     & 2004 Mar 31 & 2453095.76013 & 1x1300 & 16.416 \\

                     & 2004 Apr 01 & 2453096.74102 & 1x1300 & 15.359 \\
                     & 2004 Apr 01 & 2453096.75691 & 1x1300 & 15.369 \\

\hline
\hline
\end{tabular}
\end{table}
\clearpage
\end{document}